# Bubble formation due to capillary instability during evaporation of a porous medium

Tao Zhang[a], Rui Wu[a,b,*], C.Y. Zhao[a,b], Evangelos Tsotsas[c], Abdolreza Kharaghani[c]

[a]School of Mechanical Engineering, Shanghai Jiao Tong University, Shanghai 200240, China.

[b]Key Laboratory for Power Machinery and Engineering, Ministry of Education, Shanghai Jiao Tong University, Shanghai 200240, China

[c]Chair of Thermal Process Engineering, Otto von Guericke University, P.O. 4120, 39106 Magdeburg, Germany

* Corresponding author

E-mails:

ruiwu@sjtu.edu.cn (R. Wu)

**Abstract**

We show that during evaporation of a pore network, liquid can refill the gas occupied pores, snapping off a gas bubble, which then moves to a stable configuration. This phenomenon is induced by the capillary instability due to the wettability heterogeneity of the pore network and has a much smaller time scale as compared to the evaporation process. The capillary instability induced liquid refilling and bubble movement are explained in detail based on the analysis of the images obtained from the visualization experiment. The capillary valve effect, which hinders the movement of the gas-liquid interface and is induced by the sudden geometrical expansion between small and large pores, can be suppressed by the residual liquid in the large pore. For better understanding of the capillary instability induced gas-liquid two-phase transport during evaporation, a novel pore network model is developed, which considers not only the capillary and viscous forces but also the inertial forces that are seldom taken into account in the previous models. The pore network modeling results are in good agreement with the experimental data, demonstrating the effectiveness of the developed pore network model, which opens up a new route for better understanding of the role of inertial forces in two-phase transport in porous media.

## 1. Introduction

Evaporation from porous media is ubiquitous in nature and many industrial applications. During evaporation, the gas-liquid interfaces recede into the porous media, and the initially liquid filled pores are gradually occupied by the gas phase [1]. This gas-liquid interface movement is influenced by viscous forces [2], gravity [3], thermal gradient [4], and the pore structures [5] and wettabilities [6]. The gas-liquid interface movement determines the phase distribution in the porous media and hence the evaporation kinetics [7]. Accurate description of gas-liquid interface movement is hence of vital importance. Although much progress has been made on evaporation in porous media, both experimentally [8-10] and numerically [11,12], unveiling the dynamics of gas-liquid interface movement in porous materials is still a challenge since the porous media not only are opaque but also have complex pore structures and non-uniform wettabilities.

Visualization experiments with microfluidic pore networks composed of regular pores, which enable precise visualization of the gas-liquid interface (meniscus), is proven to be a useful approach for understanding two-phase transport phenomenon in porous media [13-20]. Based on the microfluidic visualization experiment, we observe surprisingly that during evaporation of a microfluidic pore network, liquid can refill the pores occupied by gas, snapping off a gas bubble, which then moves until a stable configuration is reached, see liquid refilling and bubble movement in fig. 1. This liquid refilling and bubble movement have a time scale about 16 seconds, much smaller than that associated with the whole evaporation process (about 30 hours). In spite of their small time scale, the liquid refilling and the bubble movement, without a doubt, influence the liquid distribution in the pore network and hence the evaporation kinetics. In this paper, we explain in detail the phenomena of liquid refilling and the bubble movement observed in our experiment. For better understanding of the observed phenomena, we also develop a novel pore network model, which takes into account not only the capillary and viscous forces but also the inertial forces that are seldom considered in the existing pore network models. The modeling and experimental results are in good agreement.

## 2. Experiments

The quasi 2D microfluidic pore network used in our experiment is fabricated by bonding a silicon wafer etched with the designed pore structures to a glass sheet. The pore network consists of $5 \times 5$ large pore bodies connected by small pore throats, fig. 1. All the pores have the same height of $h = 50$ μm. In the plane perpendicular to the height direction, the pore bodies are square and have a side length of $l = 1$ mm; the distance between the centers of two neighboring pore bodies is 2 mm; all the pore throats are rectangular with same length of $l = 1$ mm but the width, $w$, varying from 0.14 to 0.94 mm. The pore network is connected to the environment through an outlet pore of 2 mm long and 0.5 mm wide. The evaporation experiment is performed at the temperature of 26.5 ± 2 ℃ and the relative humidity of 67.3 ± 2%. The pore network, initially filled with ethanol, is horizontally placed on a plate. A camera (Nikon D810) and an inverted microscope (Olympus IX73) are used to record the movement of menisci in the pore network during evaporation. The images are analyzed to determine the liquid distribution, the curvature radii of menisci based on arc fitting method [21], the contact angle based on polynomial fitting approach [22], and the speed of the moving menisci.

The advancing and receding menisci are defined as those moving towards liquid and gas, respectively. The speed of a moving meniscus in a pore is defined as the rate of liquid flow in this pore divided by the cross-sectional area of the pore body (all the pore bodies have the same cross sectional area). Since the pore network structure is etched in the silicon wafer in our microfluidic model, we can get the contact angle (taken in the liquid phase) for the silicon surface, $\theta_s$, based on the image analysis. The contact angle hysteresis for the silicon surface is small and can be neglected. But $\theta_s$ is different in each pore due to the wettability heterogeneity of the pore network. The contact angle for the glass surface is $\theta_g = 0$ °[23].

## 3. Results and Discussion

### 3.1. Experimental results

The detailed menisci movement during liquid refilling process is presented in fig. 2a. Here, meniscus A is the advancing one, and meniscus B is the receding one. The variations of speeds and curvature radii of these two menisci are depicted in fig. 3a and 3b, respectively. In fig. 3b, $r_i$ is the curvature radius of the meniscus is the plane shown in fig. 2a. The curvature radius in the plane perpendicular to the one shown in fig. 2a is denoted $r_h$.

Before liquid refilling happens, the evaporation induced moving meniscus A is at the entrance of pore body 5. At the interface between a pore throat and a pore body, a sudden geometrical expansion exists, which can hinder the meniscus movement, i.e., the so called capillary valve effect.[14] When the triple line of meniscus A at the side wall moves to the entrance of pore body 5, the contact angle increases suddenly from $\theta_s$ to $\theta_s + 90$ °. As a result, the triple line of meniscus A at the side wall is pinned, since the triple line cannot move toward the liquid phase until the contact angle is reduced to $\theta_s$. On the other hand, since pore throats and pore bodies have the same height, the triple line at the top glass wall and the bottom silicon wall will continue moving. As a result, meniscus A changes its shape, fig. 2a, leading to the reduced contact angle. As this contact angle decreases from $\theta_s + 90°$ to $90°$, the curvature radius, $r_i$, of meniscus A decreases. But, as the contact angle further reduces from $90°$ to $\theta_s$, the curvature radius, $r_i$, increases, resulting in a lower capillary pressure, $P_c$, which in turn leads to a higher liquid pressure, $P_l$, at meniscus A, since the gas pressure, $P_g$, is assumed to be constant. The capillary pressure is $P_c = P_g - P_l = \sigma(1/r_i + 1/r_h)$, where $\sigma$ is the surface tension. The increase of the liquid pressure at meniscus A results in higher liquid pressure in the pore network, thereby leading some other menisci to be unstable and invade the gas phase, e.g., meniscus B in fig. 2a. This capillary instability results in the liquid refilling and the bubble formation and movement shown in fig. 1.

During liquid refilling, the speeds of the moving and receding menisci depend on their curvature radii, $r_i$ and $r_h$. For instance, from stages II to III shown in fig. 2a, the moving menisci speeds are always increasing, fig. 3a. This higher speed is owing to the increased curvature radius, $r_i$, of meniscus A, fig. 3b. However, the curvature radius, $r_i$, of meniscus A is smaller than that of meniscus B, stages from II to III in fig. 3b. If these two menisci have the same value of $r_h$, then liquid pressure at meniscus B will be higher than that at meniscus A, and the capillary instability induced liquid refilling and bubble formation and movement shown in fig. 1 will not occur. From this point of view, the curvature radius, $r_h$, of meniscus A should be larger than that of meniscus B. Since $r_h = h/(1+\cos\theta_s)$, the contact angel, $\theta_s$, of pore body 5 should be larger than that of pore body 1. Nevertheless, it is not easy to obtain the value of $\theta_s$ in pore bodies, since we cannot observe the moving triple line therein. However, as we will show later, pore throats have different values of $\theta_s$, indicating the wettability heterogeneity of the pore network.

At the stage of III shown in fig. 2a, meniscus A touches the side wall of the pore body 5 and the entrance of pore throat 6. Then, the inertial force drives meniscus B to continue moving toward the gas phase, which in turn pulls meniscus A to enter pore throat 6, stage IV of fig. 2a. As meniscus A enters pore throat 6 from pore boy 5, its curvature radius, $r_i$, reduces (fig. 3b), decreasing the liquid pressure at meniscus A. As a result, a sharp reduction in the menisci moving speed is observed from stages III to IV (fig. 3a), although the curvature radius, $r_i$, of meniscus B also decreases.

After meniscus A invades pore throat 6, the curvature radius, $r_i$, of meniscus B continues decreasing due to the inertial forces until it enters the pore throat 2, stage V in fig. 2a. During this process, curvature radius, $r_i$, of meniscus A in pore throat 6 remains constant, fig. 3b. The liquid pressure difference between menisci A and B increases, resulting in the higher speed of moving menisci, stages from IV to V in fig. 3a. From stages V to VI, both menisci A and B are in pore throats; hence, variations of curvature radii and speeds of moving menisci are not significant, fig. 3.

After stage VI, meniscus A invades pore body 7 from pore throat 6, stage VII in fig. 2a. The curvature radius, $r_i$, of meniscus A increases due to the capillary valve effect. The variation of $r_i$ of meniscus A during this process is not shown in fig. 3b, since the meniscus shape, as shown in stage VII of fig. 2a, is not circle due to the contaminant in pore body 7. Meniscus B is always in pore throat 2 from stages VI to VII, and the curvature radius, $r_i$, is almost constant, fig. 3b. Hence, the increase of the curvature radius, $r_i$, of meniscus A results in higher speeds of moving menisci, fig. 3a.

After stage VII, the inertial force drives meniscus A to invade pore throat 8 and meniscus B to enter pore body 3, stage VIII in fig. 2a. During this process, the curvature radius, $r_i$, of meniscus A decreases, and the curvature radius, $r_i$, of meniscus B increases, resulting in the sharp reduction in the speeds of these two menisci, fig. 3. Interestingly, although meniscus B enters the pore body 3 from pore throat 2, we find that the influence of the capillary valve effect is trivial, which is suppressed by the residual liquid in pore body 3, fig. 2b. Before meniscus B invades pore body 3, the intersectional points between pore throat 2 and pore body 3 are attached to the residual liquid. As the triple lines (at the side walls) of meniscus B move to these intersectional points between pore throat and pore body, then meniscus B will merge with menisci of the residual liquid to form a new meniscus. During this process, meniscus B keeps concave towards the gas phase, fig. 2b. By contrast, if there is no residual liquid in pore body 3, then a higher liquid pressure is needed to change meniscus B from concave to convex so as to push it into pore body 3 (more details on this invasion can be found in ref 14). Hence, it is easier for liquid to invade an empty pore body from an adjacent pore throat if the residual liquid is present in the pore body.

After the stage VIII shown in fig. 2a, meniscus B connects two pore throats, and the curvature radius, $r_i$, increases as the liquid saturation in the pore body increases. By contrast, the curvature radius, $r_i$, of meniscus A in pore throat 8 is almost constant, fig. 3b. Although meniscus B has a larger $r_i$ than meniscus A at the stage VIII, the inertial forces drive meniscus B to continue refilling of pore body 3, see stage IX in fig. 2a. After pore body 3 is completely filled by liquid, a gas bubble is then formed in the pore network, fig. 1.

### 3.2. Pore network model

For better understanding of the liquid refilling and bubble movement induced by the capillary instability, a pore network model is also developed. The inertial force is considered in the model since, as discussed above, it plays an important role during liquid refilling. The liquid flow in a pore is taken as a one-dimensional fully developed laminar flow. As shown in fig.1, the filled pore is one with the local liquid saturation $s_l > 0$ in pore throats or $s_l > s_{l,re}$ in pore bodies, for which $s_{l,re}$ is the saturation of the residual liquid. To determine $s_{l,re}$, we assume that the residual liquid in a pore body connecting the mouths of two pore throats is triangle. A partially filled pore ($s_l \leq 1$) contains menisci and has at least one adjacent empty pore, while a fully filled pore ($s_l = 1$) is surrounded by filled pores. The empty pore is the one with $s_l = 0$ in pore throats or $s_l = s_{l,re}$ in pore bodies.

The liquid flow in a filled pore throat $k$ between two filled pore bodies, $i$ and $j$, can be described as:

$$\frac{\partial \left[h w_k \rho_l \left(\frac{l_i}{2} + l_k + \frac{l_j}{2}\right) v_{l,k}\right]}{\partial t} = (P_{l,i} - P_{l,j}) h w_k - g_k \left(\frac{l_i}{2} + l_k + \frac{l_j}{2}\right) v_{l,k} \quad (1)$$

where $\rho_l$ is the liquid density, $\mu_l$ the dynamic viscosity, $v_l$ the liquid velocity, and $g_k$ is expressed as[24]:

$$g_k = \frac{\pi^4 \mu_l}{8\left[1 - \frac{2h}{\pi w_k} \tanh\left(\frac{\pi w_k}{2h}\right)\right]} \frac{w_k}{h} \quad (2)$$

The pore width and length are depicted in fig. 1. The liquid flow in a filled pore throat $k$ with $s_{l,k} < 1$ from a filled pore body $i$ to an empty pore body $j$ is depicted as:

$$\rho_l h w_k \left\{\left(\frac{l_i}{2} + s_{l,k} l_k\right) \frac{d^2(s_{l,k} l_k)}{dt^2} + \left[\frac{d(s_{l,k} l_k)}{dt}\right]^2\right\} = (P_{l,i} - P_{l,k}) h w_k - g_k \left(\frac{l_i}{2} + s_{l,k} l_k\right) \frac{d(s_{l,k} l_k)}{dt} \quad (3)$$

The liquid pressure in a partially filled pore is determined as $P_l = P_g - P_c$, for which the gas pressure $P_g$ is constant. The capillary pressure, $P_c$, for a meniscus in a pore throat $i$ is gained by the MSP method [25,26]:

$$P_c = \frac{\sigma(k_1 + 2k_2 k_4)}{w_i h - 2k_3 k_4^2} \quad (4a)$$

$$k_1 = (2h + w_i)\cos\theta_{s,i} + w_i \quad (4b)$$

$$k_2 = \pi - 3\theta_{s,i} - (2\cos\theta_{s,i} - 3\sin\theta_{s,i} + 2)\cos\theta_{s,i} \quad (4c)$$

$$k_3 = \cos^2\theta_{s,i} + \cos\theta_{s,i} - \frac{3\sin\theta_{s,i}\cos\theta_{s,i}}{2} - \frac{\pi}{2} + \frac{3\theta_{s,i}}{2} \quad (4d)$$

$$k_4 = \sqrt{\frac{k_1^2}{4k_2^2} - \frac{w_i h}{2k_3} + \frac{k_1}{2k_2}} \quad (4e)$$

The capillary pressure for a meniscus in a partially filled pore body depends on the shape of meniscus:

$$P_c = \sigma \left(\frac{1}{r_i} + \frac{1}{r_h}\right) \quad (5)$$

for which $r_h = h/(1+\cos\theta_s)$, whereas $r_i$ depends on not only the liquid saturation of the pore body but also the state (empty or filled) of the adjacent pore throat. For a partially filled pore body $i$ with only one adjacent empty pore throat (e.g., pore throat $k$ in fig. 4a),

$$r_i = \frac{w_k}{2\cos A} \quad (6a)$$

$$s_{l,i} = 1 - \frac{r_i^2 \sin(\pi - 2A) + (\pi + 2A) r_i^2}{2 l_i^2} \quad (6b)$$

where *A* is the contact angle shown in fig. 4a.

For a partially filled pore body *i* with two adjacent empty pore throats neighboring to each other (e.g., pore throats *k* an *l* in fig. 4b),

$$r_i = \frac{L_{kl}}{2\cos A} \tag{7a}$$

$$s_{l,i} = 1 - \frac{L_k L_l}{2l_i^2} - \frac{L_{kl}^2}{4l_i^2}\tan A - \frac{\pi + 2A}{2l_i^2}r_i^2 + \frac{1}{8l_i^2}(l_p - W_l)(l_p - W_k) \tag{7b}$$

where $L_k = W_k/2 + l_p/2$, $L_l = W_l/2 + l_i/2$, $L_{kl} = \sqrt{L_k^2 + L_l^2}$, and *A* the contact angle shown in fig. 4b. In fig. 4b, the meniscus in the pore body is connected to the mouths of two pore throats. As liquid in the pore body increases, the shape of the meniscus changes. When the contact angle between the wall of an empty pore throat and the meniscus reaches to $\pi - \theta_s$, the meniscus starts to enter this pore throat, see fig. 4c. If meniscus first enters pore throat *l*,

$$r_i = \frac{W_l + l_p}{2[\cos\theta_{s,l} + \cos(2A - \theta_{s,l})]} \tag{8a}$$

$$s_{l,i} = \frac{I_p - I_{t,l}}{l_p^2} \tag{8b}$$

$$I_p = \frac{1}{2}[r_i\cos\theta_{s,l} + r_i\cos(2A - \theta_{s,l})][r_i\sin(2A - \theta_{s,l}) - r_i\sin\theta_{s,l}] - \frac{1}{2}[(\pi - 2A)r_i^2 - r_i^2\sin(2A)]$$
$$+ \frac{1}{8}(W_l + l_i)(l_i - W_k) + \frac{1}{2}l_i(l_i - W_l) + \frac{1}{8}(l_i - W_l)(l_i - W_k) \tag{8c}$$

$$I_{t,l} = \frac{r_i^2\sin\beta_l}{2} + \frac{|x_1 - x_2||y_1 - y_2|}{2} - \frac{\beta_l r_i^2}{2} \tag{8d}$$

$$(x_1, y_1) = (r_i\cos\theta_{s,l}, r_i\sin\theta_{s,l}) \tag{8e}$$

$$(x_2, y_2) = \left(\sqrt{r_i^2 - y_2^2}, r_i\sin(2A - \theta_{s,l}) - (l_i + W_k)/2\right) \tag{8f}$$

$$\beta_l = \arccos\left[1 - \frac{(x_1 - x_2)^2}{2r_i^2} - \frac{(y_1 - y_2)^2}{2r_i^2}\right] \tag{8g}$$

When meniscus also enters both pore throat *k*,

$$r_i = \frac{W_l + l_i}{2[\cos\theta_{s,l} + \cos B]} \tag{9a}$$

$$s_{l,i} = \frac{I_p - I_{t,l} - I_{t,k}}{l_i^2} \tag{9b}$$

$$I_p = \frac{1}{2}(r_i\cos\theta_{s,l} - r_i\sin\theta_{s,k})(r_i\cos\theta_{s,k} - r_i\sin\theta_{s,l}) - \frac{1}{2}[(\pi - \theta_{s,l} - \theta_{s,k})r_i^2 - r_i^2\sin(\theta_{s,l} + \theta_{s,k})]$$
$$+ \frac{1}{4}(W_l + l_i)(l_i - W_k) + \frac{1}{2}l_i(l_i - W_l) + \frac{1}{8}(l_i - W_l)(l_i - W_k) \tag{9c}$$

$$I_{t,k} = \frac{r_i^2 \sin\beta_k}{2} + \frac{|x_3 - x_4||y_3 - y_4|}{2} - \frac{\beta_k r_i^2}{2} \tag{9d}$$

$$(x_3, y_3) = \left(r_i \cos\theta_{s,l} - \frac{l_i + W_l}{2}, \sqrt{r_i^2 - x_3^2}\right) \tag{9e}$$

$$(x_4, y_4) = (r_i \sin\theta_{s,k}, r_i \cos\theta_{s,k}) \tag{9f}$$

$$\beta_k = \arccos\left[1 - \frac{(x_3 - x_4)^2}{2r_i^2} - \frac{(y_3 - y_4)^2}{2r_i^2}\right] \tag{9g}$$

where $B$ is the angle between the vertical line through the center of the meniscus and the line connecting the center and the intersectional point between meniscus and vertical side wall of the pore body, as illustrated in fig. 4c. When $r_i$ of the meniscus reaches to

$$r_i = \frac{-b - \sqrt{b^2 - 4ac}}{2a} \tag{10a}$$

$$a = 1 - \cos^2\theta_{s,l} - \cos^2\theta_{s,k} \tag{10b}$$

$$b = 2W_l \cos\theta_{s,l} + (l_i + W_k)\cos\theta_{s,k} \tag{10c}$$

$$c = -W_l^2 - \left[\frac{(l_i + W_k)}{2}\right]^2 \tag{10d}$$

the meniscus completely leaves the pore body and enters pore throat $l$ and $k$.

For a partially filled pore body $i$ with two adjacent empty pore throat opposite to each other (e.g., pore throats $l$ and $m$ in fig. 4c),

$$r_i = \frac{A}{|A|} \frac{L_{lm}}{2\cos(A)} \tag{11a}$$

$$S_{l,i} = \frac{\left(l_i - \frac{W_m + W_l}{2}\right)}{2l_i} - \frac{A}{|A|l_i^2}\left(\frac{(\pi - |2A|)r_i^2}{2} - \frac{r_i^2}{2}\sin|2A|\right) + \frac{(l_i - W_k)(l_i - W_m)}{8l_i^2} + \frac{(l_i - W_l)(l_i - W_k)}{8l_i^2} \tag{11b}$$

where contact angle $A$ is positive when the meniscus is concave towards liquid, but negative when the meniscus is convex.

Based on the mass conservation law, the following equation is gained for the fully filled pore body:

$$\sum A_j v_{l,j} = 0 \tag{12}$$

where $A_j$ is the cross-sectional area of the adjacent pore throats, $v_{l,j}$ is the liquid velocity from the pore throat to the

pore body. For a partially filled pore body:

$$\sum A_j v_{l,j} + \frac{dV_{l,i}}{dt} = 0 \tag{13}$$

Where $V_{l,i}$ is the volume of liquid in the pore body.

The procedure to simulate the capillary instability induced two-phase flow in the pore network is summarized below.

(1) The liquid saturation, velocity, and pressure in each filled pore at time $t$ are given.

(2) The volume of liquid, $V_l$, in each partially filled pore with an active meniscus at time $t + \Delta t$ is determined. Here, $\Delta t = 1.7 \times 10^{-4}$ s is the step time. The liquid volume of a partially filled pore throat $j$ adjacent to a filled pore body $i$ is updated as:

$$V_{l,j}^{t+\Delta t} = hw_j l_j s_{l,j}^{t+\Delta t} \tag{14a}$$

$$s_{l,j}^{t+\Delta t} = \sqrt{\frac{2hw_j}{l_j^2 g_j} \left\{ (P_{l,i}^t - P_{l,j}^t)\Delta t + \frac{\rho_l}{g_j}\left[(P_{l,i}^t - P_{l,j}^t)hw_j - g_j s_{l,j}^t l_j v_{l,j}^t\right]\left(e^{\frac{-g_j \Delta t}{\rho_l hw_j}} - 1\right)\right\} + (s_{l,j}^t)^2} \tag{14b}$$

If the updated liquid volume in the pore throat $j$, $V_{l,j}^{t+\Delta t}$, is negative, then the liquid volume in the adjacent filled pore body $i$ is $V_{l,i}^{t+\Delta t} = V_{l,j}^{t+\Delta t} + V_{l,i}^t$, and $V_{l,j}^{t+\Delta t}$ is set to be zero. If $V_{l,j}^{t+\Delta t}$ is larger than the volume of the pore throat $j$, $V_j$, then the liquid volume in the adjacent empty or partially filled pore body $k$ is $V_{l,k}^{t+\Delta t} = V_{l,k}^t + V_{l,j}^{t+\Delta t} - V_j$, and $V_{l,j}^{t+\Delta t}$ is set to b $V_j$.

The volume of liquid in a partially filled pore body $i$ adjacent to a filled pore throat $j$ is updated as:

$$V_{l,i}^{t+\Delta t} = V_{l,i}^{t+\Delta t} + \sum A_j v_{l,j}^t \Delta t \tag{15}$$

If the updated liquid volume in the pore body $i$, $V_{l,i}^{t+\Delta t}$, is smaller than the volume of the residual liquid $V_{rel,i}$, then the liquid volume in the adjacent filled pore throat $j$ with the lowest capillary pressure is $V_{l,j}^{t+\Delta t} = V_{l,j}^t - (V_{rel,i} - V_{l,i}^{t+\Delta t})$, and $V_{l,i}^{t+\Delta t}$ is set to be $V_{rel,i}$. If $V_{l,i}^{t+\Delta t}$ is larger than the volume of pore body $i$, $V_i$, then the liquid volume in each adjacent empty or partially filled pore throat $j$ is $V_{l,j}^{t+\Delta t} = V_{l,j}^t + V_i$, and $V_{l,i}^{t+\Delta t}$ is set to be $V_i$. Here, $n_t$ is the number of adjacent empty pore throats.

Based on the volume of liquid in each pore, the liquid saturation of each filled pore at time $t + \Delta t$ is gained straightforwardly.

(3) The state of each gas-liquid interface (meniscus) is determined. A static meniscus is the one that cannot move, and an active meniscus is movable. For each partially filled pore with a static meniscus, if the pressure difference (gas pressure minus liquid pressure) across this meniscus is smaller than the capillary pressure of this pore, then the meniscus is set to be active. For each partially filled pore throat with an active meniscus and the liquid saturation being smaller than 0.1 or larger than 0.9, if the liquid flow direction is different at time $t$ and $t - \Delta t$, then the meniscus in this partially filled pore throat is labeled as static at time $t + \Delta t$ so as to avoid the numerical error. The explanation is as follows. Because of the inertial forces, a meniscus in a pore throat can enter the neighboring pore body, but the liquid pressure in the pore throat is smaller than liquid pressure in the pore body. To this end, the meniscus in the pore body may recede to the mouth of the pore throat, leading to the change of the direction of the liquid flow in the pore throat. In reality, during meniscus receding to pore throat, the three phase contact line along the side wall of the pore can be pinned at the intersection between the pore throat and pore body, and the meniscus stops moving (i.e., static).

(4) The liquid and gas clusters in the pore network are identified.

(5) The liquid pressure and velocity in each filled pore are determined by the following algorithm.

(5.1) The pressures of liquid in the partially filled pore with active menisci, $P_l^{t+\Delta t}$, are determined by the capillary pressure, see Eqs. (4) - (11). For the partially filled pore throat with static menisci, the liquid velocity, $v_l^{t+\Delta t}$, is zero. The pressures of liquid in the fully filled pore are guessed as $P_l^{*,t+\Delta t}$.

(5.2) Based on the guessed liquid pressure field, the guessed liquid velocity in each fully filled pore throat, e.g., pore throat $k$ connecting pore bodies $i$ and $j$, is determined by solving Eq. (1):

$$v_{l,k}^{*,t+\Delta t} = \frac{\left(P_{l,i}^{*,t+\Delta t} - P_{l,j}^{*,t+\Delta t}\right)\Delta t h w_k + h w_k \rho_l \left(\frac{l_i}{2} + l_k + \frac{l_j}{2}\right) v_{l,k}^t}{h w_k \rho_l \left(\frac{l_i}{2} + l_k + \frac{l_j}{2}\right) + \Delta t g_k \left(\frac{l_i}{2} + l_k + \frac{l_j}{2}\right)} \tag{16}$$

The liquid velocity in the partially filled pore throat with an active meniscus, e.g., pore throat $k$ connected to filled pore body $i$, is gained by solving Eq. (3) analytically:

$$v_{l,k}^{*,t+\Delta t} = \frac{1}{g_k l_k s_{l,k}^{t+\Delta t}} \left\{ \left(P_{l,i}^{*,t+\Delta t} - P_{l,k}^{t+\Delta t}\right) h w_k + \left[\left(P_{l,i}^{*,t+\Delta t} - P_{l,k}^{t+\Delta t}\right) - g_k l_k s_{l,k}^t v_{l,k}^t\right] e^{\frac{-g_k \Delta t}{\rho_l h w_k}} \right\} \tag{17}$$

In Eqs. (16) and (17), if the pore body is the partially filled pore with active menisci, then $P_l^{*,t+\Delta t} = P_l^{t+\Delta t}$.

(5.3) The correct liquid pressure and velocity are defined as $P_l^{t+\Delta t} = P_l^{*,t+\Delta t} + P_l^{',t+\Delta t}$ and $v_l^{t+\Delta t} = v_l^{*,t+\Delta t} + v_l^{',t+\Delta t}$, respectively. Here, $P_l'$ is the liquid pressure correction, and $v_l'$ is the liquid velocity correction. It should be noted that the correct liquid pressure and velocity also satisfy Eqs. (1) and (3), and the following equations can be gained:

$$v_{l,k}^{t+\Delta t} = \frac{\left(P_{l,i}^{t+\Delta t} - P_{l,j}^{t+\Delta t}\right)\Delta t h w_k + h w_k \rho_l \left(\frac{l_i}{2} + l_k + \frac{l_j}{2}\right) v_{l,k}^t}{h w_k \rho_l \left(\frac{l_i}{2} + l_k + \frac{l_j}{2}\right) + \Delta t g_k \left(\frac{l_i}{2} + l_k + \frac{l_j}{2}\right)} \tag{18}$$

$$v_{l,k}^{t+\Delta t} = \frac{1}{g_k l_k s_{l,k}^{t+\Delta t}} \left\{ \left(P_{l,i}^{t+\Delta t} - P_{l,k}^{t+\Delta t}\right) h w_k + \left[\left(P_{l,i}^{t+\Delta t} - P_{l,k}^{t+\Delta t}\right) - g_k l_k s_{l,k}^t v_{l,k}^t\right] e^{\frac{-g_k \Delta t}{\rho_l h w_k}} \right\} \tag{19}$$

Subtracting Eq. (16) from Eq. (18) yield the liquid velocity correction in the fully filled pore throat:

$$v_{l,k}^{',t+\Delta t} = \frac{\Delta t h w_k \left(P_{l,i}^{',t+\Delta t} - P_{l,k}^{',t+\Delta t}\right)}{h w_k \rho_l \left(\frac{l_i}{2} + l_k + \frac{l_j}{2}\right) + \Delta t g_k \left(\frac{l_i}{2} + l_k + \frac{l_j}{2}\right)} \tag{20}$$

Subtracting Eq. (17) from Eq. (19) yield the liquid velocity correction in the partially filled pore throat:

$$v_{l,k}^{',t+\Delta t} = \frac{1}{g_k l_k s_{l,k}^{t+\Delta t}} \left( h w_k + e^{\frac{-g_k \Delta t}{\rho_l h w_k}} \right) P_{l,i}^{',t+\Delta t} \tag{21}$$

(5.3) Substituting the correct liquid velocity in the filled pore throat, $v_l = v_l' + v_l^*$, into Eq. (12) yields:

$$\sum A_j \left(v_{l,k}^{',t+\Delta t} + v_{l,k}^{*,t+\Delta t}\right) = 0 \tag{22}$$

Substituting the liquid velocity correction in Eqs. (20) and (21) into Eq. (22) yields a set of linear equations for the liquid pressure correction in the each fully filled pore body. For instance, for the fully filed pore body $i$ adjacent to pore body $j$ and filled pore throat $k$, we have

$$P_{l,i}^{',t+\Delta t} = \frac{a_p - Q_p}{a_t} \tag{23a}$$

$$a_p = \sum a_{t,k} P_{l,j}^{',t+\Delta t} \tag{23b}$$

$$a_t = \sum a_{t,k} \tag{23c}$$

$$Q_p = \sum A_k v_{l,k}^{*,t+\Delta t} \tag{23d}$$

$$A_k = hw_k \tag{23e}$$

If the pore throat $k$ is fully filled, $a_{t,k} = A_k^2 \Delta t/(hw_k \rho_l l_k s_{l,k}^{t+\Delta t})$; if pore throat $k$ is partially filled with liquid saturation smaller than 1, $a_{t,k} = A_k^2[1 + exp(-g_k \Delta t/\rho_l A_k)]/(g_k l_k s_{l,k}^{t+\Delta t})$.

By solving the linear equations, the liquid pressure correction is gained for each fully filled pore body, from which the liquid velocity correction in each filled pore throat is gained based on Eqs. (20) and (21). In this way, the correct liquid pressure, $P_l^{t+\Delta t} = P_l^{*,t+\Delta t} + P_l^{',t+\Delta t}$, for each fully filled pore body and the correct liquid velocity, $v_l^{t+\Delta t} = v_l^{*,t+\Delta t} + v_l^{',t+\Delta t}$, for each filled pore throat are obtained.

(5.4) The guessed liquid pressure in each fully filled pore body is updated to equal to the correct liquid pressure gained in the step (5.3), i.e., $P_l^{*,t+\Delta t} = P_l^{t+\Delta t}$. Repeat from steps (5.2) until the prescribed convergence criteria are satisfied.

(6) Repeat the steps (2) - (5) until all menisci become static.

To simulate the observed capillary instability induced two-phase flow in the pore network, the initial liquid pressure at time $t = t_0$ in each filled pore is needed, which, however, cannot be gained experimentally. To get the initial condition, the following method is employed. Based on the visualization image, we can get the liquid velocity and volume in each filled pore at time $t = t_0 - \Delta t$. The liquid volume in each partially filled pore throat, e.g., pore throat $j$, at time $t = t_0$ is determined as $V_{l,j}^{t_0} = V_{l,j}^{t_0-\Delta t} - v_{l,j}^{t=t_0-\Delta t} \Delta t$. Then liquid volume, liquid pressure, liquid velocity in each filled pore at time $t = t_0$ can be gained based on the algorithm mentioned above, and are taken as the initial conditions.

To validate the developed model, we first simulate liquid refilling shown in fig. 2a. However, we just focus on the process from stages VI to VII, since the wettability of pore bodies are unknown. Although during this process meniscus A is in pore body 7, the adjacent pore throats 6 and 8 have very similar contact angle (40.5° and 41.7°, respectively). Hence, the contact angle of pore body 7 is taken as the averaged value of these two adjacent pore throats. The contact angle of pore throat 2 is 22.1°. The calculated speed of the moving meniscus A agrees well with the experimental data, fig. 3a.

The developed pore network model is also used to simulate the bubble movement shown in fig. 1. The modeling and experimental results are in good agreement, fig. 5. Here, the gas pressure in the bubble is assumed to be uniform and equal to the atmospheric pressure. The contact angles of pores needed in the model are shown in fig. 5a. The contact angles of pore bodies are set based on the contact angles of the neighboring pore throats. Since meniscus C shown in fig. 5a oscillates only in one pore, velocity is used instead of speed to depict its movement (fig. 5d), and flow from left to right is defined as the positive direction. It takes a relative long time for meniscus B to enter the pore body at the center of the pore network shown in fig. 5a. The main reason is that when meniscus B meets residual liquid in the pore body, its curvature radius, $r_i$, increases. This in turn hinders the meniscus movement. The good agreement between the simulations and experiments presented in fig. 5 and fig. 3a indicates the effectiveness of the present pore network model.

## 4. Conclusions

In summary, we report that during evaporation of a microfluidic pore network, liquid can refill the gas occupied pores, snapping off a gas bubble, which then moves until a stable configuration is reached. This liquid refilling and bubble formation and movement are induced by the capillary instability due to the wettability heterogeneity, and are explained in detail based on the visualization experiment. We find that the capillary valve effect, which hinders the movement of the gas-liquid interface and is induced by the sudden geometrical expansion between small and large pores, can be suppressed by the residual liquid in the large pore. In addition, a novel pore network model that considers the inertial forces is developed to simulated the experimentally observed capillary instability induced two-phase flow in the pore network. The pore network modeling results agree well with the experimental data, validating the effectiveness of the developed model, which opens up a new route for better understanding of the role of inertial forces in two-phase transport in porous media. We believe that unveiling the capillary instability induced phenomenon presented in this work can pave the way toward the fundamental understanding of the two-phase transport in porous media.

**Acknowledgments**

The authors are grateful for the support of the National Key Research and Development Program of China (No. 2018YFC1800600), and the National Natural Science Foundation of China (No. 51776122).

**REFERENCES**

(1) T.M. Shaw. 1987 Drying as an immiscible displacement process with fluid counterflow. *Physical Review Letters*. *59*, 1671.

(2) T. Metzger & E. Tsotsas. 2008 Viscous stabilization of drying front: Three-dimensional pore network simulations. *Chemical Engineering Research and Design*. **86**, 739.

(3) A.G. Yiotis, D. Salin, E.S. Tajer & Y.C. Yortsos. 2012 Drying in porous media with gravity-stabilized fronts: Experimental results. *Physical Review E*. **86**, 026310.

(4) N. Vorhauer, E. Tsotsas & M. Prat. 2018 Temperature gradient induced double stabilization of the evaporation front within a drying porous medium. *Physical Review Fluids*. **3**, 114201.

(5) S. Biswas, P. Fantinel, O. Borgman, R. Holtzman & L. Goehring. 2018 Drying and percolation in correlated porous media. *Physical Review Fluids*. **3**, 124307.

(6) O. Chapuis & M. Prat. 2007 Influence of wettability condition on slow evaporation in two-dimensional porous media. *Physical Review E.* **75**, 046311.

(7) L. Xu, S. Davies, A.B. Schofield & D.A. Weitz. 2008 Dynamics of drying in 3D porous media, *Physical Review Letters.* **101**, 094502.

(8) N. Shokri & M. Sahimi. 2012 Structure of drying fronts in three-dimensional porous media. *Physical Review E.* **85**, 066312.

(9) J. Thiery, S. Rodts, D.A. Weitz & P. Coussot. 2017 Drying regimes in homogeneous porous media from macro-to nanoscale. *Physical Review Fluids.* **2**, 074201.

(10) E. Keita, P. Faure, S. Rodts & P. Coussot. 2013 MRI evidence for a receding-front effect in drying porous meida. *Physical Review E.* **87**, 062303.

(11) M. Prat. 2011 Pore network models of drying, contact angle, and film flows. *Chemical & Engineering Technology.* **34**, 1029.

(12) T. Defraeye. 2014 Advanced computational modelling for drying processes - A review *Applied Energy.* **131**, 323.

(13) R. Wu, A. Kharaghani & E. Tsotsas. 2016 Two-phase flow with capillary valve effect in porous media. *Chemical Engineering Science.* **139**, 241.

(14) R. Wu, A. Kharaghani & E. Tsotsas. 2016 Capillary valve effect during slow drying of porous media. *International Journal of Heat and Mass Transfer.* **94**, 81.

(15) P. Panizza, H. Algaba, M. Postic, G. Raffy, L. Courbin & F. Artzner. 2018 Order-Disorder structural transitions in mazes built by evaporation drops. *Physical Review Fluids.* **121**, 078002.

(16) Y. Edery, S. Berg & D. Weitz. 2018 Surfactant variations in porous media localize capillary instabilities during haines jumps. *Physical Review Letters.* **120**, 028005.

[17] C. Odier, B. Levache, E. Santanach-Carreras & D. Bartolo. 2017 Forced imbibition in porous media: A fourfold scenario. *Physical Review Letters.* **119**, 208005.

(18) B. Zhao, C. W. MacMinn & R. Juanes. 2016 Wettability control on multiphase flow in patterned microfluidics. *Proceedings of the National Academy of Sciences.* **113**, 10251.

(19) Mingming Lv; Zhigang Liu; Can Ji; Lei Jia; Yake Jiang. Investigation of pore-scale behaviors of foam flow in polydimethylsiloxane micromodel. *Langmuir.* **2018**, *57*, 15172-15180.

(20) Panayiotis Kolliopoulos; Krystopher S. Jochem; Robert K. Lade; Jr.; Lorraine F. Francis; Satish Kumar. Capillary flow with evaporation in open rectangular microchannels. *Langmuir.* **2019**, *35*, 8131-8143.

(21) W. Tao, H. Zhong, X. Chen, Y. Selami & H. Zhao. 2018 A new fitting method for measurement of the curvature radius of a short arc with high precision. *Measurement Science and Technology.* **29**, 075014.

(22) E. Atefi, J.A. Mann, Jr. & H. Tavana. 2013 A robust polynomial fitting approach for contact angle measurements. *Langmuir.* **29**, 5677.

(23) E.F. Ethington & U.S. 1990 Interfacial contact angle measurements of water, mercury, and 20 organic liquid on quartz, calcite, biotite, and Ca-montmorillonite substrates. *Geological Survey*, 90.

(24) N. Ichikawa, K. Hosokawa & R. Maeda. 2004 Interface motion of capillary-driven flow in rectangular microchannel. *Journal of Colloid Interface Science.* **280**, 155.

(25) R.P. Mayer & R.A. Stone. 1965 Mercury porosimetry – breakthrough pressure for penetration between packed spheres. *Journal of Colloid Science.* **20**, 893.

(26) H.M. Princen. 1969 Capillary phenomena in assemblies of parallel cylinders. *Journal of Colloid and Interface Science.* **30**, 69.

**Figure Captions**

**Fig. 1** Liquid refilling and bubble formation and movement observed during evaporation of a microfluidic pore network composed of pore bodies and pore throats (the pore length and width are also illustrated). The left and right columns are the evaporation process. The middle two columns are capillary instability induced liquid refilling and bubble formation and movement processes (about 16 seconds). Gas, liquid, and solid are shown in light gray, dark gray, and black in the experimental images, respectively.

**Fig. 2** (a) Detailed evolution of menisci during capillary instability induced liquid refilling shown in fig. 1. Menisci A and B are moving, while the other menisci remain almost static. The pores that can be invaded by menisci are numbered. (b) Detailed invasion of meniscus B from pore throat 2 to empty pore body 3 with the residual liquid.

**Fig. 3** Variation of (a) menisci speeds and (b) curvature radii, $r_i$, during liquid refilling shown in fig. 2a. Menisci A and B as well as various evaporation stages are illustrated in fig. 2a. The blue line in fig. 3a is gained from the pore network simulations.

**Fig. 4** Schematic of the meniscus configuration in a partially filled body $i$ with (a) only one adjacent empty pore throat $k$; (b) (c) two adjacent empty pore throats $k$ and $l$ neighboring to each other; and (d) two adjacent empty pore throats $l$ and $m$ opposite to each other.

**Fig. 5** (a) Detailed evolution of menisci during bubble movement shown in fig. 1; top: experiment; bottom: pore network model. The pore contact angles needed in the pore network model are also shown. (b-d) Comparison of experimental and pore network modeling speeds (or velocity) of menisci A, B, and C shown in (a), respectively.

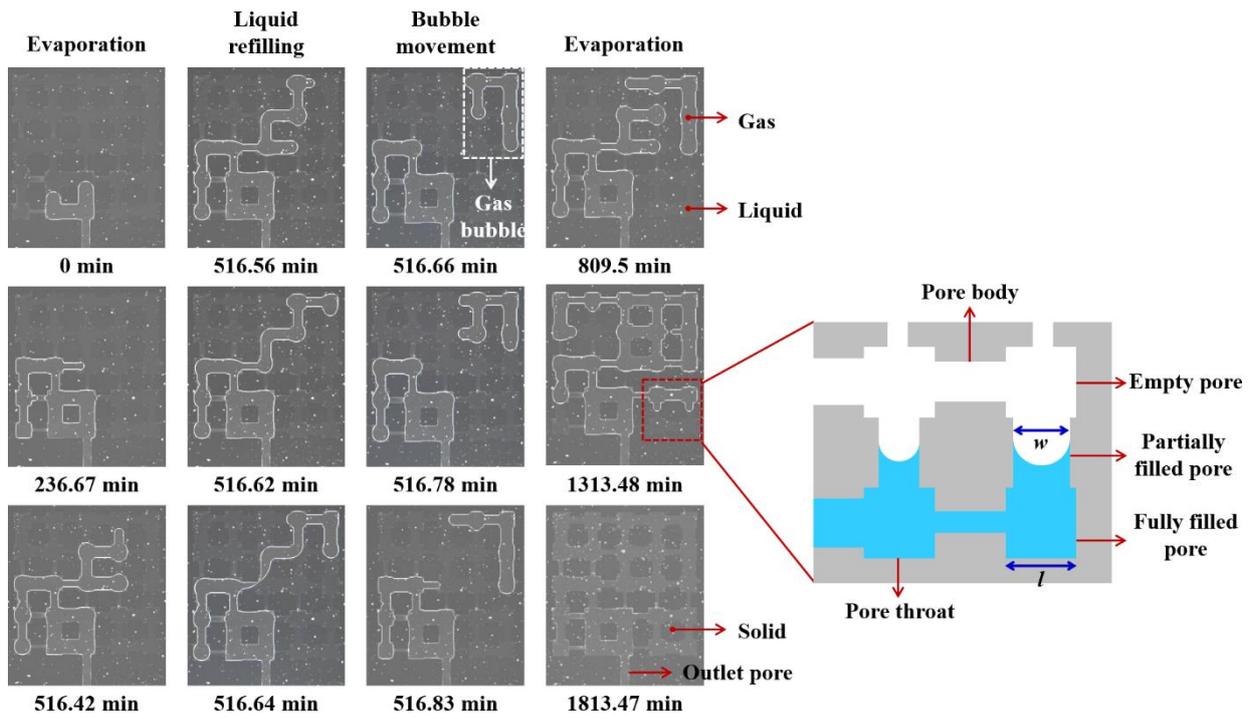

**Fig. 1**

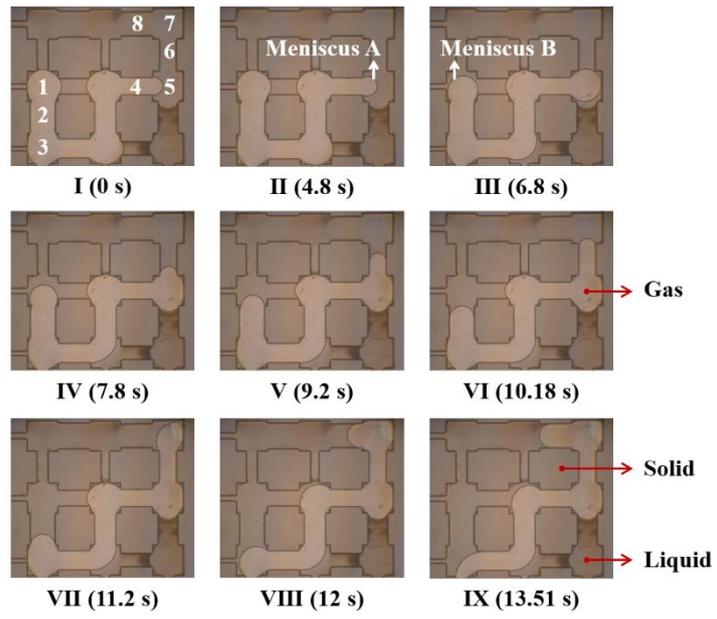

**(a)**

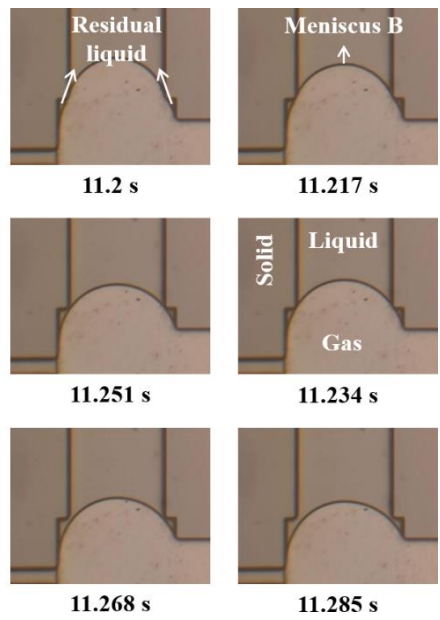

**(b)**

**Fig. 2**

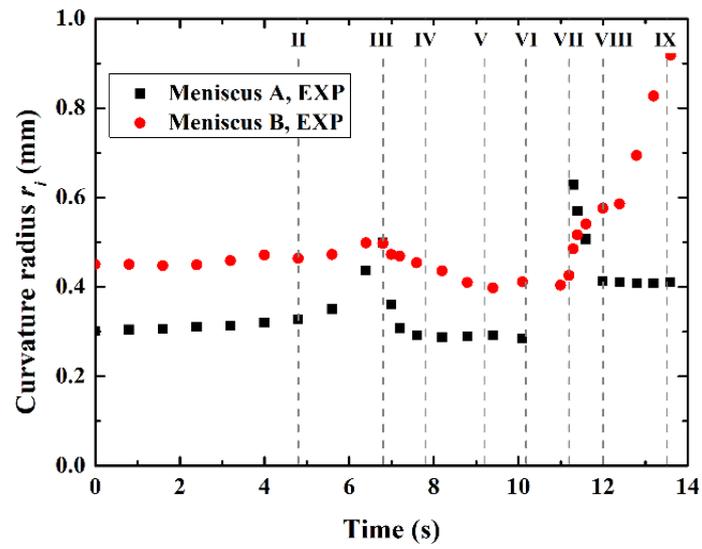

(a)

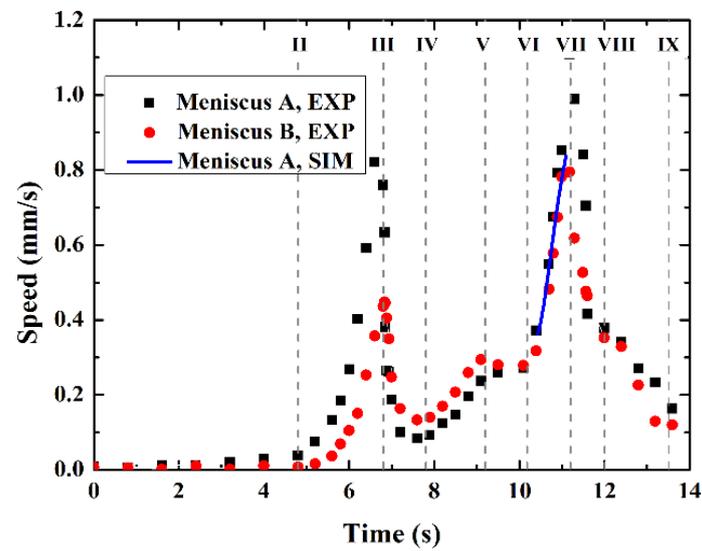

(b)

Fig. 3

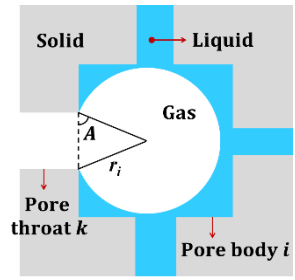

**(a)**

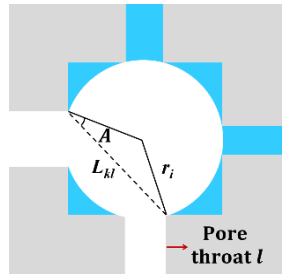

**(b)**

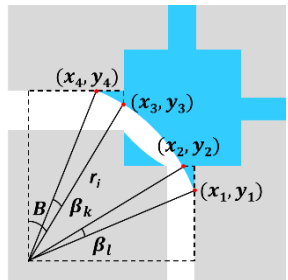

**(c)**

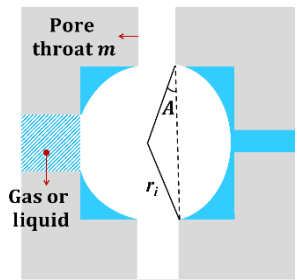

**(d)**

**Fig. 4**

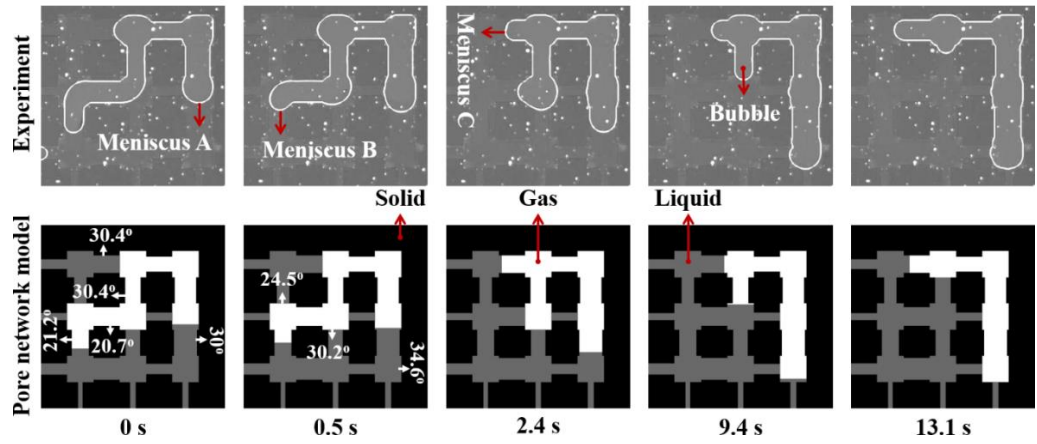

(a)

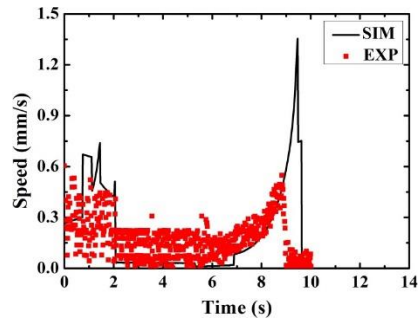

(b)

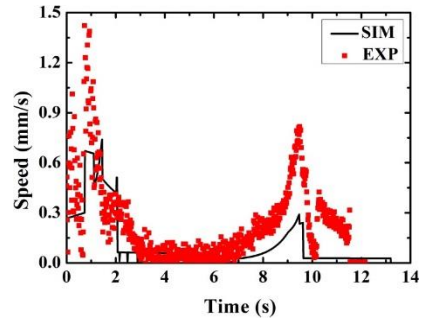

(c)

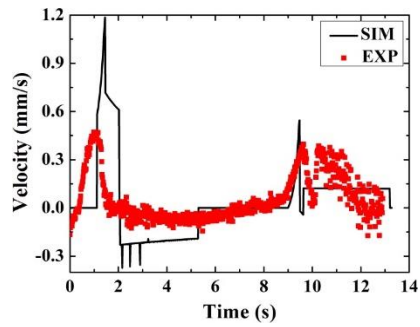

(d)

**Fig. 5**